\documentclass[%
 aip,
 amsmath,amssymb,
preprint,%
]{revtex4-1}

\usepackage{graphicx}
\usepackage{dcolumn}
\usepackage{bm}

\usepackage{amsmath,amsfonts,amssymb}
\usepackage{amsthm}
\usepackage{multirow}

\usepackage{hyperref}       
\usepackage{url}            
\usepackage{booktabs}       
\usepackage{nicefrac}       
\usepackage{microtype}      
\usepackage{fancyhdr}       
\usepackage{graphicx}        
\usepackage[utf8]{inputenc}
\usepackage[T1]{fontenc}
\usepackage{mathptmx}
\usepackage{etoolbox}

\usepackage[rgb]{xcolor}

\makeatletter
\def\@email#1#2{%
 \endgroup
 \patchcmd{\titleblock@produce}
  {\frontmatter@RRAPformat}
  {\frontmatter@RRAPformat{\produce@RRAP{*#1\href{mailto:#2}{#2}}}\frontmatter@RRAPformat}
  {}{}
}%
\makeatother
\begin{document}

\title{Predicting fluid-structure interaction with graph neural networks}
\author{Rui Gao}

\author{Rajeev K. Jaiman}%
\email{rjaiman@mail.ubc.ca}
\affiliation{ 
Department of Mechanical Engineering, University of British Columbia
}%

\date{\today}

\begin{abstract}
We present a rotation equivariant, quasi-monolithic graph neural network framework for the reduced-order modeling of fluid-structure interaction systems. With the aid of an arbitrary Lagrangian-Eulerian formulation, the system states are evolved temporally with two sub-networks. The movement of the mesh is reduced to the evolution of several coefficients via complex-valued proper orthogonal decomposition, and the prediction of these coefficients over time is handled by a single multi-layer perceptron. A finite element-inspired hypergraph neural network is employed to predict the evolution of the fluid state based on the state of the whole system. The structural state is implicitly modeled by the movement of the mesh on the solid-fluid interface; hence it makes the proposed framework quasi-monolithic. The effectiveness of the proposed framework is assessed on two prototypical fluid-structure systems, namely the flow around an elastically-mounted cylinder, and the flow around a hyperelastic plate attached to a fixed cylinder. The proposed framework tracks the interface description and provides stable and accurate system state predictions during roll-out for at least 2000 time steps, and even demonstrates some capability in self-correcting erroneous predictions. The proposed framework also enables direct calculation of the lift and drag forces using the predicted fluid and mesh states, in contrast to existing convolution-based architectures. The proposed reduced-order model via graph neural network has implications for the development of physics-based digital twins concerning moving boundaries and fluid-structure interactions.
\end{abstract}

\maketitle

\section{Introduction}
Accurate and efficient predictions and control of the spatial-temporal dynamics of fluid-structure systems are essential in various engineering disciplines. The coupling between fluid and structure can lead to complex dynamical effects such as vortex-induced vibrations and flutter/galloping \cite{Jaiman2022}. Owing to the highly nonlinear and multiscale characteristics of fluid-structure coupling, state-of-the-art computational fluid dynamics (CFD) and finite element analysis (FEA) tools based on the solution of partial differential equations are considered for engineering analysis and design. While high-fidelity CFD/FEA simulations can provide accurate prediction, the high computational cost of these simulations limits their applications for downstream design optimization and control tasks. High-fidelity CFD/FEA simulations can serve the role of a data generator for data-driven  predictions, with efficient design iterations or active control of coupled fluid-structure systems. This work is motivated by the need to make coupled fluid-structure simulations efficient for the digital twin technology, whereby multi-query analysis, optimization and control are required. 

This limitation of high-fidelity modeling based on partial differential equations has inspired the development of the so-called data-driven reduced-order modeling (ROM) techniques. By constructing low-dimensional models, ROM techniques have the potential to address the limitations of high-fidelity CFD/FEA models for efficient multi-query analysis, optimization and control tasks. Such data-driven models can be used in an offline-online manner, whereby the models can be trained to learn a low-dimensional representation of the system from the high-dimensional physical data in the offline stage, and provide efficient predictions during the online stage. Numerous data-driven techniques have been developed for low-dimensional modeling and predictive offline-online applications. Popular methods like proper orthogonal decomposition (POD) \cite{Lumley1967,Sirovich1987} and dynamic mode decomposition (DMD) \cite{Schmid2010}, as well as their variants (e.g., \cite{Towne2018,Schmidt2019,Zhang2019}), are usually based on the projection into a linear subspace. However, these methods encounter difficulty when applied to scenarios with high Reynolds numbers and convection-dominated problems, whereas one needs a significantly larger number of linear subspaces to achieve a satisfactory approximation. 

In the last decade, deep neural networks have been explored as alternatives to the aforementioned techniques. Autoencoders, as a non-linear extension of POD, have been shown to have significant advantages over POD in the compression of data \cite{Hinton2006}. Combined with convolution, which supplies the inductive bias of locality and translational equivariance, the convolutional encoder-propagator-decoder architecture has been adopted in many applications, including flow over fixed bodies in both 2D \cite{Gonzalez2018,Thuerey2020,Bukka2021,Pant2021} and 3D \cite{Gupta2022a} scenarios, fluid-structure interactions \cite{Gupta2022b,Zhang2022}, under-water noise propagation \cite{Mallik2022}, and many more. Leveraging the computational power of modern graphical processing units (GPU), the computational times needed for these convolutional neural networks are usually reduced to the level of milliseconds per step, which are orders of magnitude faster than the traditional full-order CFD simulations.

While convolutional neural networks have achieved numerous successes, they have inherent limitations and pose challenges in dealing with fluid-structure boundaries. As convolutions have to be performed on a uniform Cartesian grid, the resolution of the grid on different regions within the simulation domain has to maintain similar resolutions. Considering the flow past a bluff body as an example, the region near the body's surface is resolved by a grid of the same density as the far field. This means that one must either introduce a significant number of grids in a far-field region that does not feature much physics, or tolerate a relatively low resolution near the body's surface. The former choice of maintaining a dense grid throughout the bulk and interface requires substantially more computational resources. As a result, the balance usually aligns with the latter, inevitably leading to the loss of fine physical details and the difficulty in extracting important physical statistics like lift and drag coefficients. Despite many efforts to mitigate this issue, by forming a hybrid model with traditional machine learning techniques such as proper orthogonal decomposition \cite{Wang2018,Reddy2019}, interpolation \& projection schemes \cite{Gupta2022b}, or special mesh design \cite{Chen2021}, a completely satisfactory solution is not yet available.

Graph neural networks, recently introduced as a geometric deep learning framework, have the ability to address this difficulty, although existing applications are mostly restricted to fluid systems with fixed boundaries. As its name suggests, graph neural networks operate on graphs, which can be intuitively converted from any mesh, meaning that one can control the resolution of the different parts of the simulation domain using the strategies adopted in CFD. As a result, graph neural networks can maintain a significantly better resolution for important regions within the domain with the same mesh size compared with convolutional neural networks. With this clear advantage over convolutional neural networks, graph neural networks are recently introduced for modeling fluid flow. Applications include flow around fixed bodies like cylinder or airfoil \cite{Pfaff2020,Lino2022,Gao2023}, reacting flows \cite{Xu2021}, flow field super-resolution \cite{Belbute2020}, flow field completion \cite{He2022}, etc. Additional techniques and designs like on-the-fly graph adaptation \cite{Pfaff2020}, multi-graph with different levels of fineness \cite{Lino2022,Migus2022}, rotational equivariance \cite{Lino2022}, quadrature integration-based loss \cite{Gao2022}, and polynomial processors \cite{Xu2021} are also combined with the graph neural networks to further boost their performance. 

While these works have demonstrated the potential of graph neural networks for various fluid flow applications, no study exists for the application to fully coupled fluid-structure systems. Notably, Pfaff et al. \cite{Pfaff2020} simulated the dynamics of flags in their work, but the graph is limited to the flag itself, and the fluid flow surrounding the flag is not simulated. Li et al. \cite{Li2018} simulated the flow flushing around a free-moving rigid box within a container. However, both the flow and the box are discretized into large particles, and therefore losing the fine physical details. Most other existing graph neural network-based works on fluid flow modeling, to the best of the authors' knowledge, only focus on applications with fixed, rigid body and/or domain boundaries. 

In this work, we aim to fill this gap in the literature by modeling fluid-structure systems via graph neural network. Adopting the arbitrary Lagrangian-Eulerian formulation, we propose a rotation-equivariant quasi-monolithic graph neural network framework for modeling fluid-structure systems. The recently developed $\phi$-GNN \cite{Gao2023} is employed to predict the evolution of the fluid state based on the state of the whole fluid-structure system. Mesh and solid movements are first projected to a lower dimension via complex-valued POD (CPOD), and then the low-order POD coefficients in polar coordinates are propagated through time by a multi-layer perceptron. The solid-state is implicitly modeled by the movement of the mesh on the fluid-solid boundary. The framework is applied to two prototypical fluid-structure problems: an elastically-mounted cylinder in a uniform flow, and a hyperelastic plate attached to a fixed cylinder in a channel flow. It is demonstrated that the framework can generate stable and accurate roll-out predictions over at least thousands of time steps. More importantly, accurate lift and drag force predictions can be directly extracted from the predicted system states by simply integrating the Cauchy stress tensor at the surface of the moving solid body, which is difficult for existing convolution-based frameworks.

This article is organized as follows. In Section \ref{sec:method}, we describe the full-order fluid-structure interaction system and discuss the individual components for our proposed quasi-monolithic graph neural network framework: complex-valued proper orthogonal decomposition, multi-layer perceptrons, and $\phi$-GNN. These individual components are then assembled in Section \ref{sec:framework} to form a novel quasi-monolithic graph neural network framework. Detailed setup of the experiments on two prototypical fluid-structure interaction problems, namely the elastically-mounted cylinder system undergoing vortex-induced vibration, as well as the hyperelastic plate attached to a fixed cylinder immersed in a channel flow, are covered in Section \ref{sec:setup}. The results of these experiments are presented and discussed in Section \ref{sec:results}. We conclude the work in Section \ref{sec:conclusion}.

\section{Methodology}
\label{sec:method}
Before presenting our quasi-monolithic methodology, we first provide a brief review of the full-order representation of the fluid-structure system. Subsequently, we will introduce the individual components that will be assembled into a complete framework in Sec. \ref{sec:framework}.

\subsection{Full-order system, discretization in space and time}
\label{sec:fom}
We briefly summarize the full-order system comprising the Eulerian fluid and the Lagrangian solid, together with the traction and velocity continuity conditions at the fluid-solid interface. Under the arbitrary Lagragian-Eulerian (ALE) framework, for the coupled system between isothermal incompressible viscous fluid flow and an compressible elastic solid body, the system can be modeled as
\begin{subequations}
	\label{eq:1}
	\begin{equation}
		\label{eq:ns}
		\rho^f\frac{\partial \boldsymbol{u}^f}{\partial t} + \rho^f(\boldsymbol{u}^f-\boldsymbol{w})\cdot\nabla \boldsymbol{u}^f=\nabla\cdot\boldsymbol{\sigma}^f+\boldsymbol{b}^f \quad \text{on}\quad\boldsymbol{\Omega}^f,
	\end{equation}
	\begin{equation}
		\nabla\cdot\boldsymbol{u}^f=0 \quad \text{on}\quad\boldsymbol{\Omega}^f,
	\end{equation}
	\begin{equation}
		\label{eq:solid}
		\rho^s\frac{\partial^2\boldsymbol{\varphi}^s}{\partial t^2}=\nabla\cdot(\boldsymbol{\sigma}^s)+\rho^s\boldsymbol{b}^s\quad\text{on}\quad\boldsymbol{\Omega}^s,
	\end{equation}
\end{subequations}
with boundary conditions on the fluid-solid interface
\begin{subequations}
	\label{eq:2}
	\begin{equation}
		\boldsymbol{u}^f(t)=\boldsymbol{u}^s(t) \quad \text{on} \quad \boldsymbol{\Gamma}^{fs},
	\end{equation}
	\begin{equation}
		\label{eq:2b}
		\boldsymbol{\sigma}^s\cdot\boldsymbol{n}=\boldsymbol{\sigma}^f\cdot\boldsymbol{n}\quad\text{on}\quad\boldsymbol{\Gamma}^{fs},
	\end{equation}
\end{subequations}
along with appropriate boundary conditions on other domain boundaries. The superscripts $(\cdot)^f$ and $(\cdot)^s$ denote the state parameters for fluid and solid respectively. The fluid velocity, mesh velocity and body force are denoted by $\boldsymbol{u}^f$, $\boldsymbol{w}$ and $\boldsymbol{b}^f$ respectively within the fluid domain $\boldsymbol{\Omega}^f$, while $\boldsymbol{\varphi}^s$, $\boldsymbol{u}^s$, $\boldsymbol{\sigma}^s$ and $\boldsymbol{b}^s$ denote the displacement, the velocity, the stress and the body force for the solid body $\boldsymbol{\Omega}^s$, respectively. The solid and fluid density are denoted by $\rho^s$ and $\rho^f$ respectively. At the solid-fluid interface $\boldsymbol{\Gamma}^{fs}$, $\boldsymbol{n}$ denotes the unit outward normal vector. Assuming Newtonian fluid, the Cauchy stress tensor is written as
\begin{equation}
	\label{eq:stress}
	\boldsymbol{\sigma}^f=-p^f\boldsymbol{I}+\mu^f(\nabla\boldsymbol{u}^f+(\nabla\boldsymbol{u}^f)^T),
\end{equation}
in which $p^f$ denotes the pressure, and $\mu^f$ denotes the viscosity of the fluid. 

The governing Eqs. (\ref{eq:1}) and (\ref{eq:2}) can be re-written in the abstract dynamical form
\begin{equation}
	\label{eq:state}
	\frac{d\boldsymbol{q}}{dt}=\tilde{\boldsymbol{F}}(\boldsymbol{q}),
\end{equation}
where $\boldsymbol{q}$ denotes the state of the FSI system. In the present study, the state of the system includes the fluid state, which comprises the velocity $\boldsymbol{u}^f=(u_x^f,u_y^f)$ and the pressure $p^f$, as well as the solid state. As will be discussed later in Sec. \ref{sec:framework}, the solid state is implicitly modeled by the mesh state consisting of the mesh displacement $(\delta x, \delta y)$. The right-hand side term $\tilde{\boldsymbol{F}}$ of Eq. \ref{eq:state} represents an underlying dynamic model for the fluid-structure system described in Eqs. \ref{eq:1}-\ref{eq:stress}. If we discretize the system with a certain fixed time step $\delta t$, then Eq. \ref{eq:state} can be rewritten following forward Euler time integration:
\begin{equation}
	\label{eq:forwardEuler}
	\boldsymbol{q}(t_{n+1})-\boldsymbol{q}(t_n)=\delta\boldsymbol{q}=\boldsymbol{F}(\boldsymbol{q}(t_n)),
\end{equation}
in which $\boldsymbol{q}(t_n)$ denotes the system state at time step $t_n$, and $\boldsymbol{F}$ is the full-order system state update function in the discretized time domain.

With reduced-order modeling, one aims to construct an approximation $\hat{\boldsymbol{F}}$ to the full-order function $\boldsymbol{F}$
\begin{equation}
	\label{eq:approx}
	\boldsymbol{F}(\boldsymbol{q}(t_n))=\boldsymbol{q}(t_{n+1})-\boldsymbol{q}(t_n)\approx\delta\hat{\boldsymbol{q}}=\hat{\boldsymbol{F}}(\boldsymbol{q}(t_n),\theta),
\end{equation}
in which $\theta$ denotes the parameters of the reduced-order model $\hat{\boldsymbol{F}}$. These parameters are calculated, tuned or trained to minimize the approximation error between the reduced-order model $\hat{\boldsymbol{F}}$ and the full-order model $\boldsymbol{F}$
\begin{equation}
	\theta = \arg_{\theta}\min(\operatorname{L}(\boldsymbol{F},\hat{\boldsymbol{F}}(\theta))),
\end{equation}
in which $\operatorname{L}$ denotes a loss function that can vary for different approaches. In this work, we construct a fully data-driven reduced-order model $\hat{\boldsymbol{F}}$ with three components: complex-valued proper orthogonal decomposition (CPOD), multi-layer perceptron, and hypergraph neural network. These building blocks will be introduced in the remaining parts of this section, and assembled into a complete framework in Sec. \ref{sec:framework}.

\subsection{Proper orthogonal decomposition}
\label{sec:pod}
As will be discussed later in Sec. \ref{sec:framework}, we consider the arbitrary Lagrangian-Eulerian (ALE) formulation for the proposed quasi-monolithic graph neural network framework. With this formulation, when the solid motion is low order, it is possible to also reduce ALE mesh movement over time to a low-order approximation. In such cases, a linear projection-based reduction is suitable. We employ Proper orthogonal decomposition (POD) for such a reduction. In particular, we adopt the complex-valued variant of POD rather than the typical real-valued POD. Such a choice would ease the difficulty of enforcing rotation equivariance to the predictions generated by the network at the architecture level, as will be discussed in Sec. \ref{sec:podmlp}. Given a data matrix $\boldsymbol{X}\in\mathbb{C}^{P\times N}$, a singular value decomposition can be performed 
\begin{equation}
	\label{eq:svd}
	\boldsymbol{X}=\boldsymbol{U}\boldsymbol{\Sigma}\boldsymbol{V}^H.
\end{equation}
in which $P$ denotes the number of sampled points (typically equal to the number of vertices in a mesh), and $N$ denotes the total number of time steps available for the decomposition. Each column of the left unitary matrix $\boldsymbol{U}$ is called a mode. Assuming the singular values on the diagonal of matrix $\boldsymbol{\Sigma}$ are in decreasing order, the left-most vectors within the matrix $\boldsymbol{U}$ would be the modes capturing the most variance in the data. For a low-rank system, a reduced-order model can be constructed by projecting the data onto the first $n_r$ modes, i.e., the left-most $n_r$ vectors within the matrix $\boldsymbol{U}$,
\begin{subequations}
	\label{eq:truncate}
	\begin{equation}
		\boldsymbol{C}^{(1:n_r)} = (\boldsymbol{U}^{(1:n_r)})^H\boldsymbol{X},
	\end{equation}
	\begin{equation}
		\hat{\boldsymbol{X}} = \boldsymbol{U}^{(1:n_r)}\boldsymbol{C}^{(1:n_r)},
	\end{equation}
\end{subequations}
where $\boldsymbol{U}^{(1:n_r)}$ denotes the matrix $\boldsymbol{U}$ truncated to its left-most $n_r$ columns (i.e., retaining the first $n_r$ modes), the coefficient matrix $\boldsymbol{C}^{(1:n_r)}$ corresponds to the projected weight of the data on each of the modes, and $\hat{\boldsymbol{X}}$ denotes the reduced-order approximation of $\boldsymbol{X}$ using the first few modes. The temporal evolution of the system is then reduced using the evolution of the mode $1$ through $n_r$ coefficients over time.

\subsection{Multi-layer perceptron}
Multi-layer perceptron (MLP), alternatively called feedforward neural network, is one of the simplest modern neural networks. For an input vector $\boldsymbol{z}$, a multi-layer perceptron with $n_l-1$ hidden layers can be written as
\begin{equation}
	f(\boldsymbol{z}) = f_{n_l}\circ f_{n_l-1}\circ f_{n_l-2}\circ\cdots\circ f_2\circ f_1(\boldsymbol{z}),
\end{equation}
in which $\circ$ denotes function composition. Each layer $f_i$ is defined as
\begin{equation}
	f_i(\boldsymbol{z})=\begin{cases}
		\sigma_i(\boldsymbol{W}_i\boldsymbol{z}+\boldsymbol{b}_i) & i=1,2,\ldots,n_l-1\\
		\boldsymbol{W}_i\boldsymbol{z}+\boldsymbol{b}_i & i=n_l\\
	\end{cases},
\end{equation}
where $\boldsymbol{W}_i\boldsymbol{z}+\boldsymbol{b}_i$ is a linear transformation of the input vector $\boldsymbol{z}$. The weight matrix $\boldsymbol{W}_i$ does not have to be a square matrix, i.e., the linear transformation can project the input into a higher or lower dimension matrix. The number of rows in the weight matrix $\boldsymbol{W}_i$ is called the layer width, and the non-linear function $\sigma_i$ is usually called the activation function. In this work, multi-layer perceptrons are employed to temporally evolve the POD coefficients of the mesh movements, and are also used as the encoders, decoders, as well as update functions for the graph neural network, introduced next in Sec. \ref{sec:gnn}. 

\subsection{Graph neural network}
\label{sec:gnn}
\begin{figure}[t]
	\centering
	\includegraphics[]{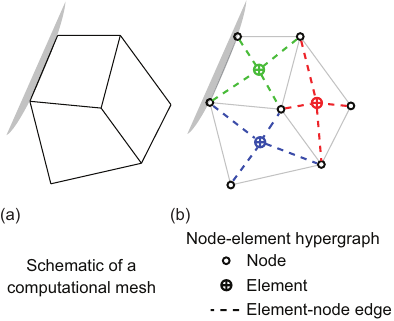}
	\caption{Conversion from a computational mesh to a node-element hypergraph. Figure modified from Fig. 2 in reference\cite{Gao2023}.}
	\label{fig:mesh2hypergraph}
\end{figure}

Graph neural networks, as discussed in the introduction, have demonstrated their advantage over convolution-based neural networks in modeling fluid systems, and therefore also utilized in this work. Specifically, we adopt the recently-developed finite element-inspired $\phi$-GNN architecture \cite{Gao2023}. Similar to several other works mentioned in the introduction, the network includes three steps: Encoding, message-passing, and decoding. Given a node-element hypergraph converted from a mesh following the illustration in Fig. \ref{fig:mesh2hypergraph}, along with the node feature $v_i$ for each node $i$, element feature $e_{\square}$ for each element $\square$, as well as element-node edge feature $e_{\square,i}$ for each element-node edge connecting node $i$ and element $\square$, the encoder step can be written as
\begin{subequations}
	\begin{equation}
		v_i\leftarrow g^v(v_i),
	\end{equation}
	\begin{equation}
		e_\square\leftarrow g^e(e_\square),
	\end{equation}
    \begin{equation}
    	e_{\square,i}\leftarrow g^{ev}(e_{\square,i}),
    \end{equation}
\end{subequations}
in which the encoder functions $g^v$, $g^e$ and $g^{ev}$ are chosen as multi-layer perceptrons. The left arrow $\leftarrow$ updates the parameter on its left-hand side by its right-hand side value.

\begin{figure}[t]
	\centering
	\includegraphics[]{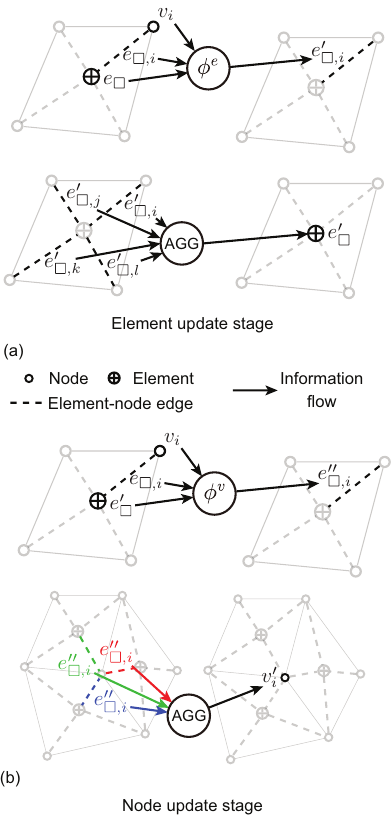}
	\caption{Schematic of (a) the element update stage and (b) the node update stage within each hypergraph message-passing layer of $\phi$-GNN. Figure modified from Fig. 3 in reference\cite{Gao2023}.}
	\label{fig:hypergraphMP}
\end{figure}

The encoded nodes, elements and element-node edge feature vectors then pass through a series of message-passing layers, with each layer consisting of an element update stage and a node update stage, plotted in Fig. \ref{fig:hypergraphMP}. For an element $\square$ connecting four nodes $i$, $j$, $k$, $l$, the element update stage is written as
\begin{subequations}
	\begin{equation}
		\label{eq:ne+mpa}
		e_{\square}\leftarrow\operatorname{AGG}_r^e\left(\phi^{e}(v_r,e_\square,e_{\square,r})\right),
	\end{equation}
	in which $\operatorname{AGG}$ denotes an aggregation function that is chosen to be the mean function in this work, and $r=i,j,k,l$ denotes the node index. The subsequent node update stage for each node $i$ can be written as
	\begin{equation}
		\label{eq:ne+mpb}
		v_i\leftarrow\operatorname{AGG}_{\square}^v\left(\phi^{v}(v_i,e_{\square_i}^\prime,e_{\square_i,i})\right).
	\end{equation}
\end{subequations}
in which $\square_i$ denotes any element that connects node $i$ with other nodes, and $e_{\square_i}^\prime$ denotes the element feature updated by the previous element update stage (Eq. \ref{eq:ne+mpa}). The element and node update functions $\phi^e$ and $\phi^v$ are not shared for different message-passing layers. The outputs at the end of the message-passing layers are decoded to produce the final outputs of the neural network. In this work, only outputs on element-node edges would be needed, and therefore only an element-node decoder $h^{ev}$ is required
\begin{equation}
	e_{\square,i}\leftarrow h^{ev}(e_{\square,i}).
\end{equation}
These outputs then pass through additional post-processing steps to obtain output vectors on nodes rather than element-node edges. Details of these treatments will be presented later in Sec. \ref{sec:gnnflow}.

\section{Quasi-monolithic graph neural network}
\label{sec:framework}
\begin{figure*}[t]
	\centering
	\includegraphics[]{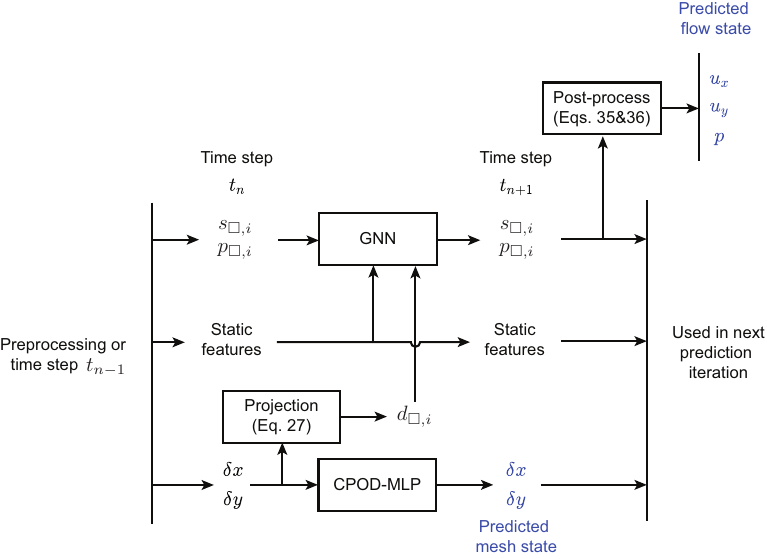}
	\caption{Schematic of the data flow within the proposed quasi-monolithic graph neural network during prediction roll-out.}
	\label{fig:dataflow}
\end{figure*}

In this section, we construct the quasi-monolithic graph neural network framework for FSI systems using the components covered in Sec. \ref{sec:method}. We choose to construct our framework under the arbitrary Lagrangian-Eulerian (ALE) formulation, reducing the complexity of lift and drag statistics extraction and also relieve the need of on-the-fly mesh adaptation. Since the mesh is body-fitting, if we only consider the movement of the solid at the solid-fluid interface, these movements can be implicitly modeled by the ALE mesh movement. With the spatial and temporal discretization discussed in Sec. \ref{sec:fom}, the purpose of the framework would be to predict the flow information $(u_x,u_y,p)$ and ALE mesh displacement information $(\delta x,\delta y)$ at all vertices of the mesh at each time step $t_n$, using flow and mesh displacement information from time step(s) before $t_n$. To serve this purpose, the proposed framework contains two sub-networks: a hypergraph neural network to temporally propagate fluid state, and a multi-layer perceptron to evolve the mesh state that has been projected to a low dimension via complex-valued POD. A schematic of the data flow during temporal roll-out is shown in Fig. \ref{fig:dataflow}. In the remaining parts of this section, we will describe the setups of the two sub-networks in detail.

\subsection{CPOD-MLP for temporal propagation of mesh}
\label{sec:podmlp}
We use a hybrid CPOD-MLP sub-network to evolve the mesh state. Rather than directly reducing the location of the vertices over time to a low dimension, we perform POD on the displacements of the vertices from their initial positions, which makes the predictions invariant to translations of the domain. For a mesh on 2D, the mesh displacements $(\delta \boldsymbol{x},\delta \boldsymbol{y})$ at each time step $t_n$ are first assembled into a complex vector
\begin{equation}
	\delta \boldsymbol{X}_{t_n}=\delta \boldsymbol{x}+\operatorname{i}\delta \boldsymbol{y},
\end{equation}
in which the non-italic $\operatorname{i}$ denotes the unit imaginary number. These complex-valued vectors in the training set are concatenated horizontally into a matrix $X$, and then decomposed and truncated following Eqs. \ref{eq:svd} and \ref{eq:truncate}, meaning that the mesh displacement vector at each time step $t_n$ can then be approximated as
\begin{equation}
	\delta \boldsymbol{X}_{t_n}\approx \boldsymbol{U}^{(1:n_r)}\boldsymbol{c}_{t_n}^{(1:n_r)}.
\end{equation}
We further convert the low-order POD coefficients $\boldsymbol{c}_{t_n}^{(1:n_r)}$ at each time step $t_n$ to polar form,
\begin{equation}
	\boldsymbol{c}_{t_n}^{(1:n_r)} = \boldsymbol{\rho}_{t_n}^{(1:n_r)}\operatorname{exp}(\operatorname{i}\boldsymbol{\eta}_{t_n}^{(1:n_r)}).
\end{equation}

\paragraph{Rotation equivariance}
Consider if the domain is rotated by any angle $\delta\eta$, then the mesh displacement vector corresponding to the rotated mesh $\delta \boldsymbol{X}_{t_n}^\prime$ can be expressed as
\begin{equation}
	\delta \boldsymbol{X}_{t_n}^\prime=\delta \boldsymbol{X}_{t_n}\operatorname{exp}(\operatorname{i}\delta\eta),
\end{equation}
which means that the coefficient corresponding to such a mesh displacement vector can be written as
\begin{equation}
	\begin{aligned}
	{\boldsymbol{c}_{t_n}^{(1:n_r)}}^\prime&=\boldsymbol{c}_{t_n}^{(1:n_r)}\operatorname{exp}(\operatorname{i}\delta\eta)\\
	(\boldsymbol{\rho}_{t_n}^{(1:n_r)})^\prime\operatorname{exp}(\operatorname{i}({\boldsymbol{\eta}_{t_n}^{(1:n_r)}}^\prime))&=\boldsymbol{\rho}_{t_n}^{(1:n_r)}\operatorname{exp}(\operatorname{i}(\boldsymbol{\eta}_{t_n}^{(1:n_r)}+\delta\eta)),
	\end{aligned}
\end{equation}
meaning that the rotation of the field only changes the arguments of the coefficients by $\delta\eta$ but would not affect the moduli. In addition, while the arguments are changed by the rotation, their increment over every time step is not affected,
\begin{equation}
	\begin{aligned}
	{\boldsymbol{\eta}_{t_{n+1}}^{(1:n_r)}}^\prime-{\boldsymbol{\eta}_{t_n}^{(1:n_r)}}^\prime&=\boldsymbol{\eta}_{t_{n+1}}^{(1:n_r)}+\delta\eta-\boldsymbol{\eta}_{t_n}^{(1:n_r)}-\delta\eta\\
	&=\boldsymbol{\eta}_{t_{n+1}}^{(1:n_r)}-\boldsymbol{\eta}_{t_n}^{(1:n_r)}.
	\end{aligned}
\end{equation}
Based on these observations, the predictions of the network would be equivariant to domain rotation if all the inputs and outputs of the network are the moduli of the coefficients as well as the increments of their arguments over time steps. It should be noted that complex-valued POD is performed without centering the data first, which is also necessary to achieve rotation equivariance with this approach.

\paragraph{Time stepping, post-processing}
In implementation, we also predict the increments of moduli rather than moduli themselves, and also use information from multiple history time steps to reduce the difficulty of predictions. The eventual inputs and outputs for the multi-layer perceptron with $n_h$ input history time steps can be written as
\begin{equation}
	\begin{bmatrix}
		\delta\hat{\boldsymbol{\rho}}_{t_{n+1}}^{(1:n_r)}\\
		\delta\hat{\boldsymbol{\eta}}_{t_{n+1}}^{(1:n_r)}
	\end{bmatrix}
\approx\operatorname{MLP}\left(
\begin{bmatrix}
	\boldsymbol{\rho}_{t_{n}}^{(1:n_r)}\\
	\boldsymbol{\rho}_{t_{n-1}}^{(1:n_r)}\\
	\vdots\\
	\boldsymbol{\rho}_{t_{n-n_h+1}}^{(1:n_r)}\\
	\delta\boldsymbol{\eta}_{t_{n}}^{(1:n_r)}\\
	\delta\boldsymbol{\eta}_{t_{n-1}}^{(1:n_r)}\\
	\vdots\\
	\delta\boldsymbol{\eta}_{t_{n-n_h+1}}^{(1:n_r)}
\end{bmatrix}\right),
\end{equation}
where $\delta\boldsymbol{\rho}_{t_{n}}^{(1:n_r)}=\boldsymbol{\rho}_{t_{n}}^{(1:n_r)}-\boldsymbol{\rho}_{t_{n-1}}^{(1:n_r)}$, and $\delta\boldsymbol{\eta}_{t_{n}}^{(1:n_r)}=\boldsymbol{\eta}_{t_{n}}^{(1:n_r)}-\boldsymbol{\eta}_{t_{n-1}}^{(1:n_r)}$.
The predicted moduli and arguments of the coefficients at time step $n+1$ can then be calculated from their values and the network predictions following forward Euler scheme
\begin{equation}
	\begin{bmatrix}
	    \hat{\boldsymbol{\rho}}_{t_{n+1}}^{(1:n_r)}\\
	    \hat{\boldsymbol{\eta}}_{t_{n+1}}^{(1:n_r)}
	\end{bmatrix}
=
\begin{bmatrix}
	\hat{\boldsymbol{\rho}}_{t_{n}}^{(1:n_r)}\\
	\hat{\boldsymbol{\eta}}_{t_{n}}^{(1:n_r)}
\end{bmatrix}
+
\begin{bmatrix}
	\delta\hat{\boldsymbol{\rho}}_{t_{n+1}}^{(1:n_r)}\\
	\delta\hat{\boldsymbol{\eta}}_{t_{n+1}}^{(1:n_r)}
\end{bmatrix}.
\end{equation}
The predicted low-order POD coefficients are then multiplied with the corresponding modes to recover the predicted mesh displacement vector at each time step
\begin{equation}
	\delta\hat{\boldsymbol{X}_{t_n}}=\boldsymbol{U}^{(1:n_r)}\hat{\boldsymbol{\rho}}_{t_{n}}^{(1:n_r)}\operatorname{exp}(\operatorname{i}\hat{\boldsymbol{\eta}}_{t_{n}}^{(1:n_r)}),
\end{equation}
with the real and imaginary parts of this vector corresponding to the $x$ and $y$ component of the mesh displacement respectively
\begin{equation}
	\begin{aligned}
	\delta \hat{\boldsymbol{x}}_{t_n}=\operatorname{real}(\delta\hat{\boldsymbol{X}_{t_n}}),\\
	\delta \hat{\boldsymbol{y}}_{t_n}=\operatorname{imag}(\delta\hat{\boldsymbol{X}_{t_n}}).
	\end{aligned}
\end{equation}

\subsection{Graph neural network for temporal propagation of flow state}
\label{sec:gnnflow}
We use the $\phi$-GNN discussed in Sec. \ref{sec:gnn} for the temporal propagation of flow state. In this subsection, we lay out the pre-processing and post-processing steps needed for the network to work within the framework. It should be noted that some of these steps are identical to the ones discussed in reference\cite{Gao2023}, but we still include them for completeness.

\paragraph{Geometry and domain features}
We transform the initial mesh coordinates to features that are translation and rotation invariant. Given an element $\square$ connecting nodes $i$, $j$, $k$, and $l$, we have the center coordinate of the element 
\begin{equation}
	(x_\square,y_\square)=(\overline{x_r},\overline{y_r}),
\end{equation}
with the overline $\overline{(\cdot)}$ denoting the mean over $r=i,j,k,l$. The local coordinates of the nodes relative to the element center can then be written as
\begin{equation}
	(x_{\square,r},y_{\square,r})=(x_r-\overline{x_r},y_r-\overline{y_r}).
\end{equation}
The length of these local coordinate vectors $L_{\square,r}=\sqrt{x_{\square,r}^2+y_{\square,r}^2}$ are taken as a geometry feature of the element, along with the angles at the four corners of the element $\theta_{\square,r}$ and the area of the element $S_\square$. These features fully constrain the shape and size of the element. Apart from the mesh shape, we transform the boundary conditions into an one-hot node feature vector $\boldsymbol{\gamma}_i$ for each node $i$ following the practice in reference\cite{Pfaff2020}.

\paragraph{ALE mesh feature}
With the geometry features ready, we proceed to process the remaining mesh and fluid information available. For the ALE mesh feature, we project the mesh displacement $(\delta x_i,\delta y_i)$ onto the directions of the unit local coordinate vectors $\boldsymbol{x}_{\square,i}$ of the node,
\begin{equation}
	\label{eq:projmesh}
	d_{\square,i} = \begin{bmatrix}
		\delta x_i&\delta y_i
	\end{bmatrix} \boldsymbol{x}_{\square,i} = \frac{1}{L_{\square,i}}\begin{bmatrix}
		\delta x_i&\delta y_i
	\end{bmatrix}\begin{bmatrix}
		x_{\square,i}\\
		y_{\square,i}
	\end{bmatrix}.
\end{equation}
The weights of these projections $d_{\square,i}$ are translation invariant and rotation equivariant, and therefore suitable to be used as features.

\paragraph{Fluid flow features}
The fluid flow information available are the flow velocity $(u_x,u_y)$ and the pressure $p$ at each node, as well as the Reynolds number $\operatorname{Re}$. The flow velocity at each node, similar to that of the mesh displacement, is projected onto the direction of the unit local coordinate vectors $\boldsymbol{x}_{\square,i}$ of the node, 
\begin{equation}
	\label{eq:projvel}
	s_{\square,i} = \begin{bmatrix}
		u_{x,i}&u_{y,i}
	\end{bmatrix} \boldsymbol{x}_{\square,i} = \frac{1}{L_{\square,i}}\begin{bmatrix}
		u_{x,i}&u_{y,i}
	\end{bmatrix}\begin{bmatrix}
		x_{\square,i}\\
		y_{\square,i}
	\end{bmatrix},
\end{equation}
With the geometry features being translation and rotation invariant, the scalar pressure is automatically translation and rotation invariant, and therefore do not need to be projected like the mesh movement or flow velocity.

\paragraph{Feature attachment}
\label{sec:preprocess}
We proceed to attach the transformed system information as features. With a body-fitting mesh, the boundary condition vector $\boldsymbol{\gamma}_i$ is attached as the node feature $v_i$, 
\begin{equation}
	v_i=\boldsymbol{\gamma}_i.
\end{equation}
The area of each element $S_\square$ is naturally the element feature. Rather than directly using the area values, their natural logarithm is used since the area can vary by orders of magnitude between large and small elements. With additional augmentation, we have the element feature vector
\begin{equation}
	e_\square=\begin{bmatrix}
		\ln S_\square\\
		-\ln S_\square
	\end{bmatrix}.
\end{equation}
The pressure value $p_i$ on each node $i$ is gathered to all the element-node edges connected to the node,
\begin{equation}
	p_{\square_i,i}=p_i.
\end{equation}
We further non-dimensionalize the pressure value and then take its signed square root before concatenating it with other available information to form the element-node edge feature vector
\begin{equation}
	\label{eq:nefeatcyl}
	e_{\square,i}=\begin{bmatrix}
		s_{\square,i}&
		\operatorname{sgn}(p_{\square,i})\sqrt{\frac{2|p_{\square,i}|}{\rho^fU_\infty^2}}&
		d_{\square,i}&
		\ln L_{\square,i}&
		\cos \theta_{\square,i}
	\end{bmatrix}^T,
\end{equation}
in which $\operatorname{sgn}(\cdot)$ denotes the sign function, and $U_\infty$ denotes the average inlet velocity or free-stream velocity. When modeling for a range of Reynolds numbers, the Reynolds number is explicitly supplied, giving
\begin{equation}
	\label{eq:nefeataf}
	e_{\square,i}=\begin{bmatrix}
		s_{\square,i}&
		\operatorname{sgn}(p_{\square,i})\sqrt{\frac{2|p_{\square,i}|}{\rho^fU_\infty^2}}&
		d_{\square,i}&
		Re&
		\ln L_{\square,i}&
		\cos \theta_{\square,i}
	\end{bmatrix}^T.
\end{equation}

\paragraph{Time stepping}
With the forward Euler scheme (Eq. \ref{eq:forwardEuler}), the network iteratively predicts the increment of the feature vectors between neighboring time steps. Specifically, the projected weights of the velocity and the gathered pressure values on the element-node edge feature vectors are updated in each time step by the graph neural network
\begin{equation}
	\label{eq:timestep}
	\begin{aligned}
		e_{\square,i}^{t_n+1}&=e_{\square,i}^{t_n}+\begin{bmatrix}
			s_{\square,i}^{t_n+1}-s_{\square,i}^{t_n}\\
			\operatorname{sgn}(p_{\square,i}^{t_n+1})\sqrt{\frac{2\left|p_{\square,i}^{t_n+1}\right|}{\rho^fU_\infty^2}}-\operatorname{sgn}(p_{\square,i}^{t_n})\sqrt{\frac{2\left|p_{\square,i}^{t_n}\right|}{\rho^fU_\infty^2}}\\
			\boldsymbol{0}_{4\times1}
		\end{bmatrix}\\
		&\approx e_{\square,i}^{t_n}+\begin{bmatrix}
			\widehat{\boldsymbol{G}}(v_i,e_{\square,i}^{t_n},e_\square)\\
			\boldsymbol{0}_{4\times1}
		\end{bmatrix},
	\end{aligned}
\end{equation}
with $\widehat{\boldsymbol{G}}$ denoting the $\phi$-GNN described in Sec. \ref{sec:gnn}.

\paragraph{Post-processing}
The predicted element-node features eventually go through the post-processing step to be converted back to velocity $(u_x,u_y)$ and pressure $p$ on each node. The scalar pressure value at a certain node can be retrieved by calculating the mean value across all the element-node edges connecting with that node,
\begin{equation}
	p_i^{t_n}=\operatorname{MEAN}_\square(p_{\square_i,i}^{t_n}).
\end{equation}
For the vector quantities of velocity and mesh displacement, we reverse the projection process described in Eq. \ref{eq:projmesh} and \ref{eq:projvel} through a Moore-Penrose pseudo-inverse \cite{Lino2022,Gao2023}
\begin{equation}
\label{eq:invgeoproj}
\begin{aligned}
\boldsymbol{u}_i^{t_n}=&\begin{bmatrix}
	u_{x,i}^{t_n}\\
	u_{y,i}^{t_n}\\
\end{bmatrix}\\
=&\left((\boldsymbol{1}_{2\times1}(\boldsymbol{s}_{\square_i,i}^{t_n})^T)\odot((\boldsymbol{X}_{\square_i,i}\boldsymbol{X}_{\square_i,i}^T)^{-1}\boldsymbol{X}_{\square_i,i})\right)\boldsymbol{1}_{n_{\square_i,i}\times1},
\end{aligned}
\end{equation}
in which the symbol $\odot$ denotes the Hadamard product, $n_{\square_i,i}$ denotes the number of element-node edges connected to node $i$, $\boldsymbol{s}_{\square_i,i}$ is a $n_{\square_i,i}\times1$ vector containing the projection weights $s_{\square_i,i}$ for all the element-node edges connected to node $i$, and $\boldsymbol{X}_{\square_i,i}$
is another $2\times n_{\square_i,i}$ matrix containing the unit local coordinate vectors $\boldsymbol{x}_{\square,i}$ for all the element-node edges connected to node $i$.

When a node $i$ only is only connected by one element (e.g., at the concave edges of domain boundaries), Eq. \ref{eq:invgeoproj} cannot be solved due to the problem being under-constrained, so we simply set $\boldsymbol{s}_{\square_i,i}^n\equiv0$ and $\widehat{\boldsymbol{u}}_i^n\equiv0$ for the predictions on those nodes.

\section{Experiments}
\label{sec:setup}
In this section, we evaluate the framework constructed in Sec. \ref{sec:framework}, focusing on its capability to learn the temporal evolution of the FSI system and make accurate temporal roll-out predictions. We consider two representative FSI systems, namely an elastically-mounted cylinder in a uniform flow, and a hyperelastic plate attached to a fixed cylinder in a channel flow. In the remaining parts of this section, we will first introduce the generation of the ground truth training and testing data sets, followed by the setup of the model and how it is trained. We then discuss the metrics to be used for the evaluation. The results will be presented and discussed in the subsequent Sec. \ref{sec:results}.

\subsection{Ground truth data generation}
In this subsection, we discuss how these data are generated, as well as their subsequent organization into training and testing data sets.

\subsubsection{Elastically-mounted cylinder} 
\begin{figure}[t]
	\centering
	\includegraphics[]{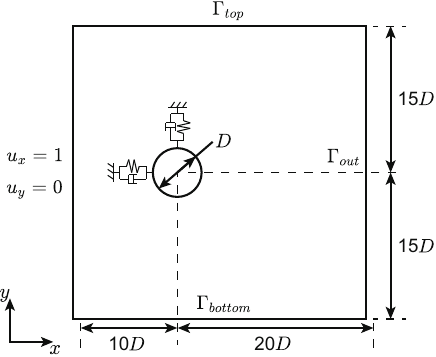}
	\caption{Schematic of the simulation domain used in the elastically-mounted cylinder case}
	\label{fig:cfddomain}
\end{figure}

\begin{figure}[t]
	\centering
	\includegraphics[]{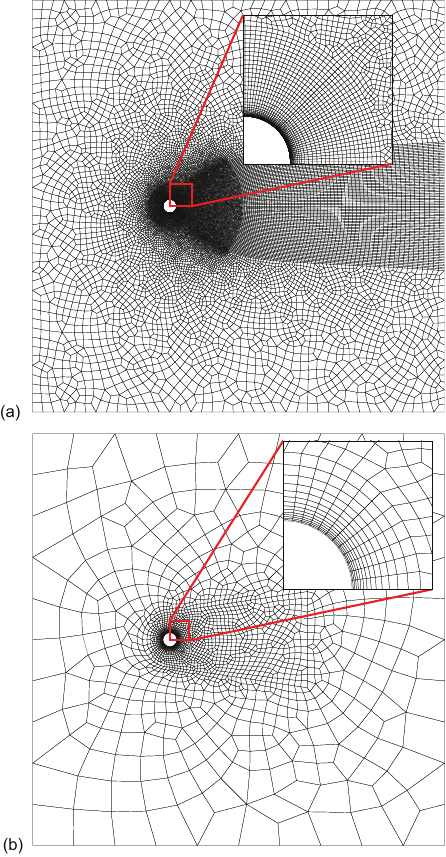}
	\caption{Schematic of the meshes used in the elastically-mounted cylinder case: (a) Mesh used for the ground truth CFD simulation, and (b) Mesh used for the proposed quasi-monolithic framework}
	\label{fig:cylmesh}
\end{figure}

For the elastically-mounted cylinder case, assuming that the cylinder is rigid and irrotational, the governing equation for the solid movement (replacing Eq. \ref{eq:solid}) can be written as
\begin{equation}
	\boldsymbol{m}^s\frac{\partial\boldsymbol{u}^s}{\partial t}+\boldsymbol{c}^s\boldsymbol{u}^s+\boldsymbol{k}^s(\boldsymbol{\varphi}^s(t)-\boldsymbol{\varphi}^s(0))=\boldsymbol{F}^s+\boldsymbol{b}^s \quad\text{on}\quad\boldsymbol{\Omega}^s,
\end{equation}
with boundary condition on the surface of cylinder replacing Eq. \ref{eq:2b}
\begin{equation}
	\boldsymbol{F}^s(t)=-\int_{\boldsymbol{\Gamma}^{fs}(t)}{\boldsymbol{\sigma}^f\cdot\boldsymbol{n}d\boldsymbol{\Gamma}}.
\end{equation}
in which $\boldsymbol{m}^s$, $\boldsymbol{c}^s$ and $\boldsymbol{k}^s$ denotes the mass, damping, and stiffness matrices respectively, whilst $\boldsymbol{\varphi}^s$, $\boldsymbol{u}^s$, $\boldsymbol{F}^s$ and $\boldsymbol{b}^s$ denote the displacement, the velocity, the traction force and the body force for the solid body $\boldsymbol{\Omega}^s$, respectively. The vibration of a cylinder is governed by four key non-dimensional parameters, namely mass-ratio $\left(m^*\right)$, Reynolds number $\left(Re\right)$, reduced velocity $\left(U_r \right)$ and critical damping ratio $(\zeta)$
defined as
\begin{equation}
	m^{*}= \frac{m^s}{m^f}, \ \ \ \ Re = \frac{\rho^{f} U_{\infty} D}{\mu^\mathrm{f}}, \ \  \ \ U_r = \frac{U_{\infty}}{f_{n} D}, \ \
	\ \  \zeta = \frac{c^s}{2\sqrt{k^s m^s}},
	\label{eq:DampingRatio}
\end{equation}
where $m^s$ is the mass of the body, $c^s$ and $k^s$ are the damping and stiffness 
coefficients, respectively for an equivalent mass-spring-damper system of a vibrating structure, $U_{\infty}$ and $D$ denote the free-stream speed and the diameter of cylinder, respectively. The natural frequency of the body is given by $f_n=(1/2\pi) \sqrt{k^s/m^s}$, the mass of displaced fluid by the structure is $m^f = \rho^f\pi D^2/4$ for a cylinder cross-section. For the current work, we consider the mass ratio $m^*=10$, the reduced velocity $U_r=5$ and the damping coefficient $c^s=0$, with a series of Reynolds numbers within the range of $[100,300]$. In particular, we sample with the Reynolds number interval of 5, leading to a total of 41 simulations at Reynolds numbers $100,105,\ldots,300$. In these simulations, the full-order systems are solved using Petrov-Galerkin finite elements and a semi-discrete time-stepping scheme \cite{Jaiman2016}. The domain and mesh for the ground-truth simulation are plotted in Fig. \ref{fig:cfddomain} and \ref{fig:cylmesh}a. The inlet has a uniform flow $u_x=U_\infty=1, u_y=0$, the outlet boundary $\Gamma_{out}$ is traction free, and a slip-wall boundary condition is implemented at top and bottom boundaries $\Gamma_{top}$ and $\Gamma_{bottom}$. 
The distance between the side walls is $30D$, and the distances to the upstream and the downstream boundaries are $10D$ and $20D$ respectively. The simulation results at $Re=200$ are validated with the reference values from Li et al. \cite{Li2016}.

For each simulation, a total of 2499 time steps continuous time steps are sampled from stable condition with a time step $\Delta t^*=0.04$, which is equivalent to the ground truth time step of the full-order simulation. These flow data are then interpolated onto a coarser mesh (Fig. \ref{fig:cylmesh}b) before being used for the training and testing of the neural network. The ground truth ALE mesh movement of the coarser mesh at each time step is calculated from the ground truth displacement $\boldsymbol{\varphi}^m=(\boldsymbol{\delta x},\boldsymbol{\delta y})$ of the cylinder by modeling the mesh as an elastic material in equilibrium, which is given by
\begin{equation}
	\nabla\cdot\left((1+k_m)[\nabla\boldsymbol{\varphi}^m+(\nabla\boldsymbol{\varphi}^m)^T+(\nabla\cdot\boldsymbol{\varphi}^m)\boldsymbol{I}]\right)=0,
\end{equation}
with boundary conditions
\begin{subequations}
	\begin{equation}
		\boldsymbol{\varphi}^m(t)=\varphi^s(t)-\varphi^s(0)\quad\text{on}\quad\boldsymbol{\Gamma}^{fs},
	\end{equation}
	\begin{equation}
		\boldsymbol{\varphi}^m=0\quad\text{on}\quad\partial\boldsymbol{\Omega}^f(0)\backslash\boldsymbol{\Gamma}^{fs},
	\end{equation}
\end{subequations}
where $\partial\boldsymbol{\Omega}^f(0)\backslash\boldsymbol{\Gamma}^{fs}$ denotes the boundary of the fluid domain excluding fluid-solid interface, and $k_m$ is the local element level mesh stiffness parameter \cite{Jaiman2022,Masud1997} which we choose as 0 for this case for simplicity. The fluid state parameters $u_x,u_y,p$ are then interpolated from the CFD mesh (Fig. \ref{fig:cylmesh}a) to the coarser mesh (Fig. \ref{fig:cylmesh}b) via the Clough-Tocher 2D interpolation scheme available in SciPy package \cite{Virtanen2020}.

The interpolated flow data and calculated mesh movement data are then organized into training and testing data sets. The flow data at Reynolds number $Re=100,110,\ldots,290,300$ are used as training and cross-validation data sets, whilst the flow data at Reynolds number $Re=105,115,\ldots,295$ are used as test data sets. For the training data set at each Reynolds number, we take 2048 continuous samples starting from time step $n_h+1$, with $n_h=7$ being the number of history time steps used for the mesh prediction sub-network. For the mesh prediction sub-network, the first training sample in the training data set at each Reynolds number can be written as
\begin{equation}
	\label{eq:meshpredsample}
		\text{Input: }\begin{bmatrix}
			\boldsymbol{\rho}_2^{(1:n_r)}\\
			\vdots\\
			\boldsymbol{\rho}_{t_{n_h}+1}^{(1:n_r)}\\
			\delta\boldsymbol{\eta}_2^{(1:n_r)}\\
			\vdots\\
			\delta\boldsymbol{\eta}_{t_{n_h}+1}^{(1:n_r)}
		\end{bmatrix},\quad\quad
		\text{Output: }\begin{bmatrix}
			\boldsymbol{\rho}_{t_{n_h}+2}^{(1:n_r)}-\boldsymbol{\rho}_{t_{n_h}+1}^{(1:n_r)}\\
			\delta\boldsymbol{\eta}_{t_{n_h}+2}^{(1:n_r)}
		\end{bmatrix},
\end{equation}
with the number of low-order POD modes used $n_r=2$. For the flow prediction network, the first training sample in the training data set at each Reynolds number can be written as
\begin{equation}
	\label{eq:flowpredsample}
	\begin{aligned}
		\text{Input: }&\text{Hypergraph with feature vectors }e_{\square,i}^{n_h+1}, e_{\square}, v_i,\\
		\text{Output: }&\text{Element-node feature vector }\\
		&\quad\begin{bmatrix}
			s_{\square,i}^{t_{n_h}+2}-s_{\square,i}^{t_{n_h}+1}\\
			\operatorname{sgn}(p_{\square,i}^{t_{n_h}+2})\sqrt{\frac{2\left|p_{\square,i}^{t_{n_h}+2}\right|}{\rho^fU_\infty^2}}-\operatorname{sgn}(p_{\square,i}^{t_{n_h}+1})\sqrt{\frac{2\left|p_{\square,i}^{t_{n_h}+1}\right|}{\rho^fU_\infty^2}}
		\end{bmatrix}.
	\end{aligned}
\end{equation}
For the testing data set at each Reynolds number, we take 2001 continuous time steps starting from time step $n_h+1$. The inputs to the network are constructed using the sample at time step $n_h+1$ following Eqs. \ref{eq:meshpredsample} and \ref{eq:flowpredsample}. The network will be applied iteratively to these inputs to generate roll-out predictions for the 2000 time steps after time step $n_h+1$, and these predictions will be compared with the ground truth values. The cross-validation is performed on the training data sets following the approach tests are performed on the test data sets.

\subsubsection{Hyperelastic plate attached to a fixed cylinder}
For the case of a hyperelastic plate attached to a fixed cylinder, the plate is attached to the end of the cylinder, and such a structure is placed in a 2D laminar channel flow. For the hyper-elastic plate, the material is specified by giving the first Piola–Kirchhoff stress tensor $\boldsymbol{\sigma}^s$
\begin{equation}
	\boldsymbol{\sigma}^s=\lambda^s(\operatorname{tr}\boldsymbol{E})\boldsymbol{F}+2\mu^s\boldsymbol{F}\boldsymbol{E},
\end{equation}
where $\boldsymbol{F}=I+\nabla\boldsymbol{\varphi}^s$ is the deformation gradient tensor, $\boldsymbol{E}=(\boldsymbol{F}^T\boldsymbol{F}-\boldsymbol{I})/2$ is the Green-Lagrange strain tensor, and $\lambda^s$ and $\mu^s$ are the Lam\'e coefficients.

\begin{figure}[t]
	\centering
	\includegraphics[]{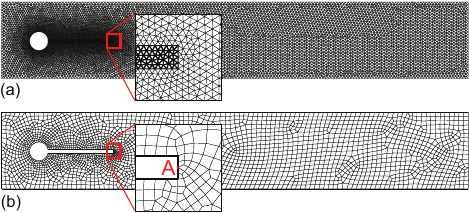}
	\caption{Schematic of the meshes used for the Turek-Hron case: (a) Mesh used for the ground truth CFD simulation, and (b) Mesh used for the proposed quasi-monolithic framework}
	\label{fig:turekmesh}
\end{figure}

In this work, we specifically use the setup of the FSI2 test within the well-known Turek-Hron benchmark \cite{turek2006proposal}, and we will therefore refer to the system as the Turek-Hron case in short in the remaining parts of this paper. We refer the readers to the original paper for the details of the case setup. The ground truth data used in this work are generated via the open-source toolbox redbKIT \cite{quarteroni2015reduced} with the computational mesh plotted in Fig. \ref{fig:turekmesh}a. Different from the case of the elastically-mounted cylinder, the ALE mesh movement is handled by the solid-extension mesh moving technique \cite{stein2004automatic} with the stiffening power $\chi=1.5$ for the whole fluid domain. 

A total of 4499 continuous time steps are sampled from a stable condition with time step $\Delta t=2.5\times10^{-3}$, equivalent to the time step of the ground truth simulation. Similar to the case of the elastically-mounted cylinder, we also interpolate the flow data onto a coarser mesh (plotted in Fig. \ref{fig:turekmesh}b). The ground truth ALE mesh movement of the coarser mesh, however, is directly interpolated from that of the CFD mesh, rather than being calculated again. The interpolated flow and mesh movement data are then separated into training and testing data sets. The training data set consists of 2048 continuous samples starting from time step $n_h+1$ with $n_h=3$ and $n_r=3$, and the samples are constructed in the same way as described in Eqs. \ref{eq:meshpredsample} and \ref{eq:flowpredsample}. The testing data set consists of 2001 continuous time steps starting from time step $n_h+2051$. 

\subsection{Model and training details}
With the training and testing data sets available, we proceed to set up and train the model. The implementation is completed with the PyTorch package \cite{paszke2019}. For the multi-layer perceptron used for the mesh displacement prediction, three hidden layers are used with layer width 512. For the graph neural network used in the fluid state prediction, the encoders, the decoder, and the node and element update functions within each network layer are all multi-layer perceptrons with two hidden layers. Sinusoidal activation function \cite{Sitzmann2020} is adopted along with the associated initialization scheme for all multi-layer perceptrons used in this work. Following the choice in reference\cite{Pfaff2020}, we fix the layer width at 128 for the multi-layer perceptrons used in the graph neural network and use 15 message-passing layers, since both of which seem to be also appropriate for our cases in the balance between accuracy and inference speed during preliminary tests. The mean aggregation function is used for both the node and element update stages. No normalization layer is used to avoid instability in training. The hypergraph message-passing layers are implemented using the gather-scatter scheme similar to the practice in PyTorch Geometric package \cite{Fey2019}. 

The two sub-networks are trained separately on a single Nvidia RTX 3090 GPU with a central processing unit (CPU) being AMD Ryzen 9 5900 @ 3 GHz $\times$ 12 cores. An Adam optimizer \cite{Kingma2014} with $\beta_1=0.9$ and $\beta_2=0.999$ is used to train the networks, with the highest learning rate being $10^{-4}$ and the lowest learning rate being $10^{-6}$. The mesh prediction sub-network is trained with batch size 256, while the flow prediction sub-network is trained with batch size 4. The networks are trained for a total of 200 epochs with the exception of the flow prediction sub-network for the elastically-mounted cylinder case, which is only trained for 50 epochs. Table \ref{tab:train} includes the details on the number of warmup and training epochs used. The learning rate is increased per training iteration during the warmup stage and decreased per epoch during the normal training stage following the learning rate scheme in reference \cite{Gao2023}. 

A smooth $L_1$ loss function \cite{girshick2015fast} is used in the training,
\begin{equation}
	\begin{aligned}
		\operatorname{L}_i(\psi_i,\hat{\psi}_i) = 
		\begin{cases}
			(\psi_i-\hat{\psi}_i)^2/2\beta,&\text{if }\lvert\psi_i-\hat{\psi}_i\rvert<\beta\\
			\lvert\psi_i-\hat{\psi}_i\rvert-\beta/2,  &\text{if }\lvert\psi_i-\hat{\psi}_i\rvert\geq\beta\\
		\end{cases}
	\end{aligned}
\end{equation}
with the non-negative control parameter $\beta$ adapted on-the-fly in every training iteration \cite{Gao2023}
\begin{equation}
	\beta^2\leftarrow (1-\frac{1}{N_{b}})\beta^2+\frac{1}{N_{b}}\min\{\beta^2,\operatorname{MSE}(\psi_{batch},\hat{\psi}_{batch})\}.
\end{equation}
Here $\psi$ denotes the ground truth output for the batch, $\hat{\psi}$ denotes the predicted output from the neural network, the subscript $(\cdot)_i$ denotes entry-by-entry calculation, $N_b$ denotes the number of training iterations per epoch, and $\operatorname{MSE}$ denotes the mean-square error. Training noise is added during the training of the flow prediction networks following the training noise scheme in reference \cite{Gao2023}. The standard deviation of the noise is set to be $0.1$ for the elastically-mounted cylinder case and $0.05$ for the Turek-Hron case, with over-correction factor $\omega=1.2$.

\begin{table}[]
	\caption{\label{tab:train} Partial details of the neural network training process}
	\begin{ruledtabular}
	\begin{tabular}{cccccc}
		Case & Sub-network & \begin{tabular}[c]{@{}c@{}}Warmup\\ epochs\end{tabular} & \begin{tabular}[c]{@{}c@{}}Training\\ epochs\end{tabular} & \begin{tabular}[c]{@{}c@{}}Batch\\ size\end{tabular} & \begin{tabular}[c]{@{}c@{}}Learning rate\\ min/max\end{tabular} \\
		\hline
		\multirow{2}{*}{Cylinder} & Mesh & 0 & 200 & 256 & \multirow{4}{*}{$10^{-6}/10^{-4}$} \\
		& Flow & 5 & 45 & 4 &  \\
		\multirow{2}{*}{Turek-Hron} & Mesh & 0 & 200 & 256 &  \\
		& Flow & 10 & 190 & 4 & 
	\end{tabular}
\end{ruledtabular}
\end{table}

\subsection{Evaluation criteria}
\label{sec:criteria}
After training the sub-networks, we test the framework on the test data sets to evaluate its performance. We use several qualitative and quantitative metrics to facilitate the evaluation. For the predicted mesh states, we plot out the ground truth versus the predicted low-order mode coefficients. We also plot out the ground truth versus the predicted trajectory of a certain point of interest for both cases: The cylinder center for the elastically-mounted cylinder case, and the point $A$ marked on Fig. \ref{fig:turekmesh}b for the Turek-Hron case. 

For the predicted fluid states, we adopt the coefficient of determination $R^2$ to qualitatively compare the ground truth versus predicted flow states for all the test data sets, which is defined as
\begin{equation}
	R^2=1-\frac{\|\boldsymbol{q}-\hat{\boldsymbol{q}}\|_2^2}{\|\boldsymbol{q}-\bar{\boldsymbol{q}}\|_2^2},
\end{equation}
for any system state parameter $\boldsymbol{q}$ and its prediction $\hat{\boldsymbol{q}}$. A coefficient of determination close to 1 means that the error of the prediction is low. In addition, we also extract the lift coefficient $C_l$ and drag coefficient $C_d$ on the solid body from the predicted system state. The calculation is performed by integrating the Cauchy stress tensor $\boldsymbol{\sigma}^f$ (defined in Eq. \ref{eq:stress}) from the first layer of elements away from the solid body,
\begin{subequations}
	\label{eq:liftdrag}
	\begin{equation}
		C_l=\frac{1}{\frac{1}{2}\rho^f(U_\infty)^2D}\int_{\boldsymbol{\Gamma}^{fs}}(\boldsymbol{\sigma}^f\cdot \boldsymbol{n})\cdot \boldsymbol{n}_y d\boldsymbol{\Gamma},
	\end{equation}
	\begin{equation}
		C_d=\frac{1}{\frac{1}{2}\rho^f(U_\infty)^2D}\int_{\boldsymbol{\Gamma}^{fs}}(\boldsymbol{\sigma}^f\cdot \boldsymbol{n})\cdot \boldsymbol{n}_x d\boldsymbol{\Gamma},
	\end{equation}
\end{subequations}
where the characteristic length $D$ is the diameter of the cylinder for both cases in this work. It should be noted that this way of calculation is different from our earlier work \cite{Gupta2022b}, as only pressure lift and drag coefficients were considered in that work, whilst in this work we also include the contribution from viscous forces. It is also different from works like reference\cite{Zhang2022}, as they did not calculate the lift and drag coefficients from the predicted flow and mesh states, but rather model and evolve them separately. For additional qualitative comparison, we plot out the ground truth versus the predicted non-dimensionalized pressure fields for several test data sets at several particular prediction time steps.

To further verify the desired invariance and equivariance properties, we also test the network on a transformed test data set obtained by applying a random shift and a rotation of $\pi/4$ to the test data set of the Turek-Hron case, and plot out the predicted versus the ground truth pressure fields for qualitative visual verification.

\section{Results and discussion}
\label{sec:results}
\subsection{Flow around elastically-mounted cylinder}
Starting from the first $n_h=7$ time steps in the test data sets for the elastically-mounted cylinder case, we run the trained framework to generate roll-out predictions of the system over the next 2000 time steps. We present and discuss these results in this subsection following the evaluation criteria discussed in Sec. \ref{sec:criteria}.

\begin{figure}[t]
	\centering
	\includegraphics[]{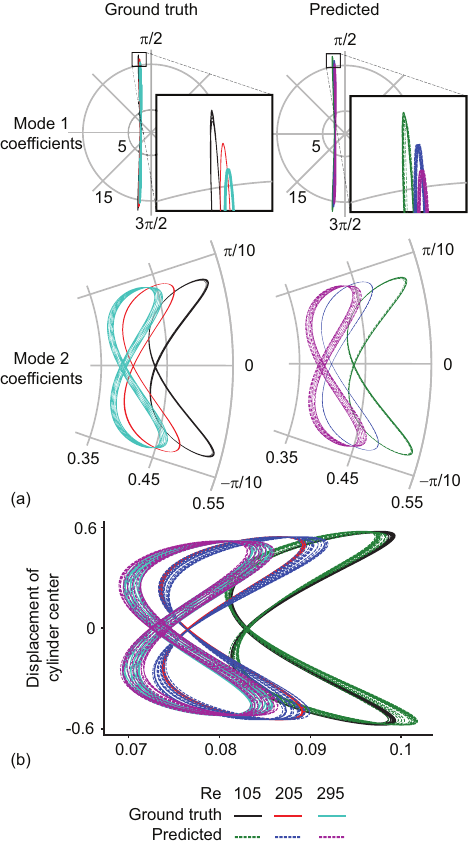}
	\caption{(a) The ground truth versus the predicted mode 1 and 2 coefficients, as well as (b) the ground truth versus the predicted cylinder center displacements for test data sets with Reynolds numbers 105, 205, and 295. Trajectories over 2000 prediction roll-out steps are plotted.}
	\label{fig:cylModes}
\end{figure}

\paragraph{Mesh state predictions}
The first two POD modes are used for the mesh state predictions. We plot out the ground truth versus the predicted mode coefficients on the polar coordinates for test data sets with Reynolds number $Re=105, 205, 295$ in Fig. \ref{fig:cylModes}a. It is clear that the mesh prediction network produces stable and reasonably accurate predictions of the mode coefficients over time. We further plot out the ground truth versus the predicted cylinder center over time for those three test data sets in Fig. \ref{fig:cylModes}b.

\begin{figure}[t]
	\centering
	\includegraphics[]{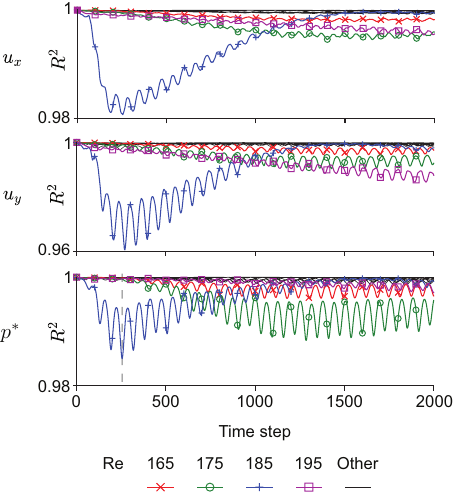}
	\caption{The coefficients of determination $R^2$ of the predicted flow states $u_x$, $u_y$ and $p^*$ for all the test data sets, with the values corresponding to test data sets at Reynolds numbers $165,175,185,195$ marked out. The vertical dashed line in the $R^2$ plot for $p^*$ marks out the time step 255 where the $R^2$ value is at its lowest. The corresponding non-dimensionalized pressure field is plotted in Fig. \ref{fig:pstarcylworst}.}
	\label{fig:r2cyl}
\end{figure}

\begin{figure}[t]
	\centering
	\includegraphics[]{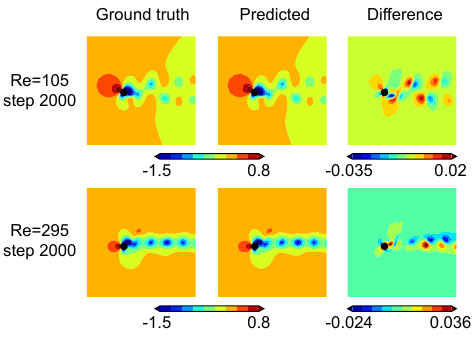}
	\caption{The ground truth versus the predicted pressure fields for test data sets with Reynolds numbers 105 and 295, at prediction roll-out time step 2000, zoomed in to the region near the cylinder.}
	\label{fig:pstarcyl}
\end{figure}

\paragraph{Flow state predictions}
Fig. \ref{fig:r2cyl} shows the coefficients of determination of the predicted flow states $u_x$, $u_y$ and $p^*$ for all the test data sets in the flow around elastically-mounted cylinder case. The coefficient of determination values are all close to 1, indicating reasonably accurate predictions of the flow states during the 2000 steps of prediction roll-out. For further qualitative comparisons, we plot out the ground truth versus the predicted flow states at prediction time step 2000 for test data sets with Reynolds number $Re=105$ and $Re=295$ in Fig. \ref{fig:pstarcyl}. Both of them only show slight errors in the predicted wake. 

\begin{figure}[t]
	\centering
	\includegraphics[]{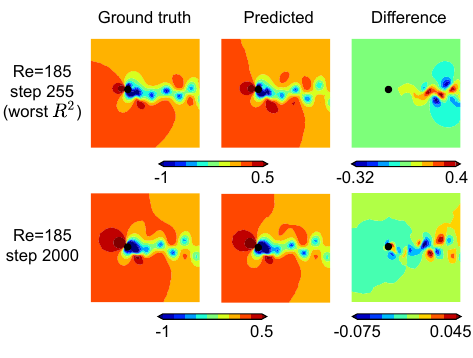}
	\caption{The ground truth versus the predicted pressure fields for test data set at Reynolds numbers 185,  for roll-out time steps 255 and 2000, zoomed in to the region near the cylinder.}
	\label{fig:pstarcylworst}
\end{figure}

\paragraph{Transition between 2D and 3D}
As marked out in Fig. \ref{fig:r2cyl} by different colors and markers, the network's predictions for test data sets $Re=165,175,185$, and $195$ are less accurate than the others. Such a range ($Re\in(160,200)$) of less accurate predictions overlaps with the Reynolds number range where the flow physics change significantly (transition from 2D to 3D), meaning that the predictions might be improved by increasing the Reynolds number sampling density for the training data sets within this Reynolds number range. Nevertheless, the network is still able to generate stabilized predictions with the existing training data sets, and, for certain cases, the prediction error seems to be gradually corrected over the prediction roll-out. We demonstrate this behavior using the test data set at Reynolds number 185, plotting the pressure field at prediction time step 255 (worst $R^2$) and 2000. At time step 255, the coefficient of determination of the predicted pressure field is at its lowest value throughout the prediction roll-out, with the prediction error concentrating in the wake region -- the fluctuation strength of the flow at the wake is under-predicted. However, such an error is gradually self-corrected during the prediction roll-out, as shown in the gradually increasing coefficient of determination after step 255 in Fig. \ref{fig:pstarcylworst}. Eventually at prediction roll-out time step 2000, the prediction error is mostly corrected, as is plotted in Fig. \ref{fig:pstarcylworst}.

\begin{figure}[t]
	\centering
	\includegraphics[]{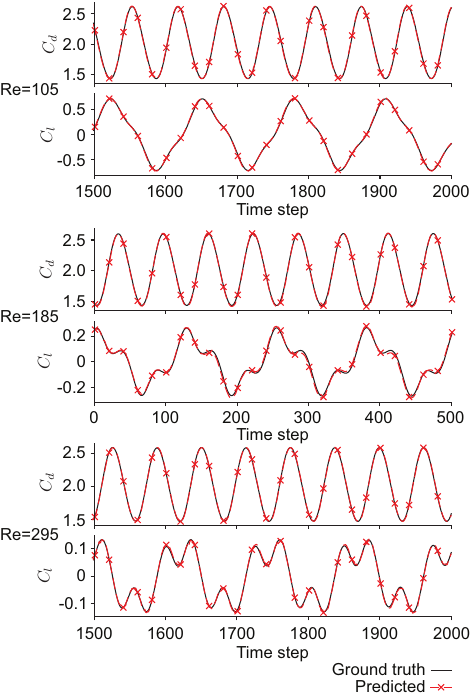}
	\caption{Lift and drag coefficients calculated from the predicted versus the ground truth system states for test data sets at Reynolds numbers 105, 185 and 295 for the elastically-mounted cylinder case. Notably, we plot these coefficients for the range between prediction roll-out time step 1501 and 2000 for $Re=105$ and $Re=295$, while plotting between time step 1 and 500 for $Re=185$.}
	\label{fig:liftdragcyl}
\end{figure}

\paragraph{Lift and drag predictions}
We further plot out the lift and drag coefficients calculated from the predicted versus ground truth mesh and fluid states for test data sets at several different Reynolds numbers in Fig. \ref{fig:liftdragcyl}. It is evident that the lift and drag coefficients calculated from the predicted system states are accurate. The lift and drag coefficient values are still accurate for the test data set at $Re=185$ around time step 255, since the prediction error at the wake region would not affect the lift and drag calculation at the surface of the cylinder.

\subsection{Hyperelastic plate attached to a fixed cylinder}
Starting from the first $n_h=3$ time steps in the test data set for the Turek-Hron case, we again evaluate the framework by generating roll-out predictions for 2000 time steps. We present these results in this subsection.

\begin{figure}[t]
	\centering
	\includegraphics[]{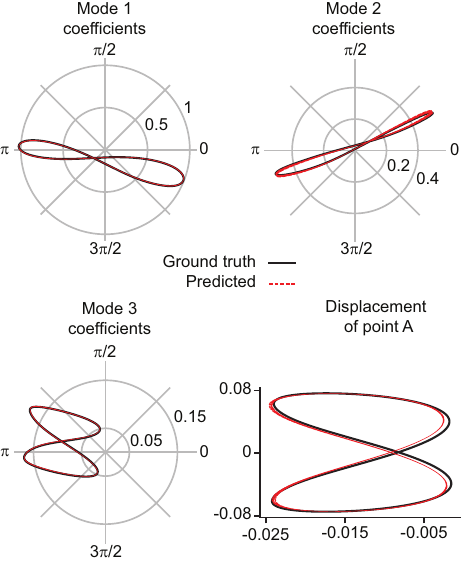}
	\caption{The ground truth versus the predicted mode 1 to 3 coefficients, as well as the ground truth versus predicted displacements of point A marked out in Fig. \ref{fig:turekmesh}, for the Turek-Hron case.}
	\label{fig:turekModes}
\end{figure}

\begin{figure}[t]
	\centering
	\includegraphics[]{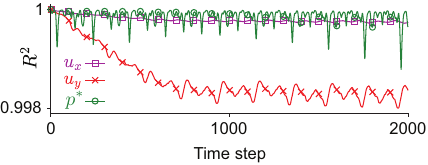}
	\caption{The coefficient of determination for the predicted fluid states over the prediction roll-out for the test data set of the Turek-Hron case.}
	\label{fig:r2turek}
\end{figure}

\begin{figure}[t]
	\centering
	\includegraphics[]{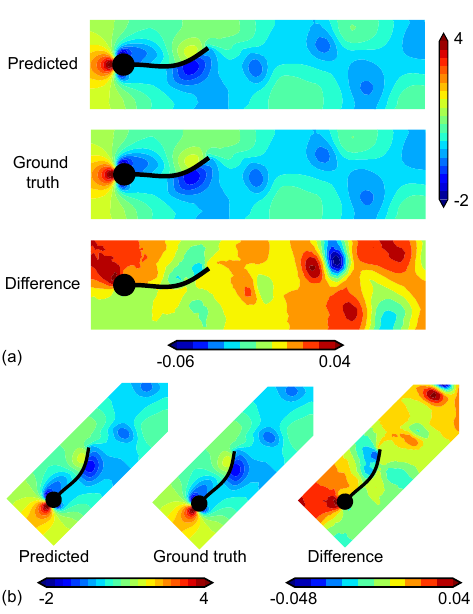}
	\caption{The predicted versus the ground truth pressure fields at prediction roll-out time step 2000 for the test data set of the Turek-Hron case. (a). Roll-out test at normal setup (b). Roll-out test with the test data set rotated by $\pi/4$ and shifted randomly. The training data sets are not rotated. The far wake is not plotted.}
	\label{fig:pstarturek}
\end{figure}

\begin{figure}[t]
	\centering
	\includegraphics[]{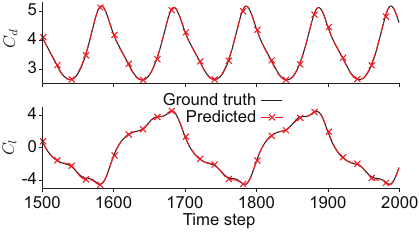}
	\caption{Lift and drag coefficients calculated from the predicted versus the ground truth system states between time step 1501 and 2000 for the test data set of the Turek-Hron case.}
	\label{fig:liftdragturek}
\end{figure}

\paragraph{System state predictions, lift and drag coefficients}
The first three POD modes are used for the mesh state predictions. We plot the ground truth versus predicted mode coefficients as well as the ground truth versus predicted displacement of point A in Fig. \ref{fig:turekModes}. For the fluid states, we plot out the coefficient of determination for $u_x$, $u_y$ and $p^*$ in Fig. \ref{fig:r2turek}, and the predicted versus ground truth pressure fields at prediction time step 2000 in Fig. \ref{fig:pstarturek}a. We further plot out the lift and drag coefficients for the whole solid body in Fig. \ref{fig:liftdragturek}. It is evident that stable and accurate system state predictions can be obtained, and that accurate lift and drag statistics can be calculated using these predicted system states. 

\paragraph{Invariance and equivariance properties}
To verify the desired invariance and equivariance properties discussed in Sec. \ref{sec:framework}, we apply a random shift and a rotation of $\pi/4$ on the test data set, and apply the network trained on the original training data set on such a transformed test data set. The predicted versus the ground truth pressure fields at prediction time step 2000 is plotted in Fig. \ref{fig:pstarturek}b, demonstrating that the desired invariance and equivariance properties are satisfied.

\subsection{Discussions}
Whilst the quasi-monolithic framework has demonstrated its efficiency, some might wonder why a monolithic approach with a conventional graph neural network is not adopted. There are two reasons for this. As mentioned earlier in Sec. \ref{sec:pod}, the first and foremost reason to consider modeling the ALE mesh movement separately is that such movements are inherently low-order when the solid motion is low-order, therefore a linear projection-based reduction can help reduce the dimension and the associated computational overhead. The flow, on the other hand, is usually highly non-linear when the Reynolds number is not low, and therefore not suitable for the same treatment. One might argue that by simply adding the increment of $\delta x$ and $\delta y$ to the neural network output, the additional computational cost is minimal. However, during preliminary tests, we noticed that such a treatment, surprisingly, makes the training of the graph neural network much more difficult. Such an observation suggests that making the additional predictions of $\delta x$ and $\delta y$ is actually costly -- a significant proportion of the expressive power of the graph neural network is expended on making these predictions, which justifies the idea of separately modeling the mesh movement. 

\begin{figure}[t]
	\centering
	\includegraphics[]{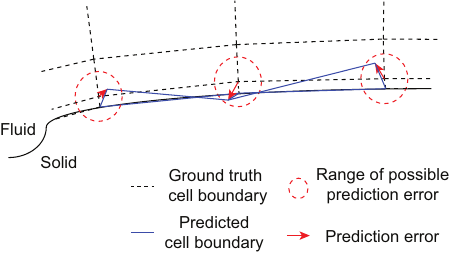}
	\caption{Illustration of the possible cell distortion with near-zero or negative Jacobian in the layer of mesh near the solid-fluid interface due to minor prediction errors of vertex location.}
	\label{fig:meshzigzag}
\end{figure}

The other reason for adopting an order reduction for the mesh movements is that the requirements for prediction accuracy will be very high when a graph neural network is directly used for the prediction of mesh movements, especially at a boundary layer where the mesh is very thin in direction perpendicular to the interface. As demonstrated by Fig. \ref{fig:meshzigzag}, a slight prediction error in the mesh displacement may lead to a distorted cell with near-zero or negative Jacobian, creating difficulty in the subsequent extraction of lift and drag statistics as well as the flow fields themselves. In contrast, when the mesh movement is reduced down to a low order approximation, the predicted mesh locations will be a linear combination of several mesh movement patterns, and this issue is much more unlikely to occur.

The utilization of POD helps to reduce computational overhead and circumvent the possible issues with Jacobian. On the other hand, it leads to an additional limitation on the application of the proposed framework, since POD only works with a certain mesh setup. Whilst the utilization of ALE formulation means that online mesh adaptation is usually not necessary, the usage of POD still limits the generalization towards other mesh configurations of the same problem or other problems. It might be possible to mitigate this limitation by adopting meshfree order reduction methods. It should also be noted that we are not modeling the solid except for its boundary at the solid-fluid interface. In application, the stress distribution within the solid might also be needed. In these cases, one could perform order reductions of the solid and fluid mesh movements together, and recover the stress from the predicted solid movements in post-processing steps.

\section{Conclusion}
\label{sec:conclusion}
We presented a rotation equivariant, quasi-monolithic graph neural network-based framework for data-driven reduced-order modeling of fluid-structure interaction systems. Adopting the arbitrary Lagrangian-Eulerian formulation, the proposed framework relies on a multi-layer perceptron for predicting low-order complex-valued POD coefficients to evolve mesh displacements, and a hypergraph neural network for predicting fluid states from the current system state. The framework is applied to the system of an elastically mounted cylinder in uniform flow, as well as the system of a hyperelastic plate attached to a fixed cylinder in channel flow. Starting from the system state at a time step along with a short history of mesh movement over the previous time steps, the framework can produce stable and accurate predictions of the system state, up to at least 2000 future time steps, with some capability to self-correct erroneous predictions. Accurate fluid and mesh state predictions on a non-uniform mesh, refined near the solid-fluid interface, enable direct and accurate extraction of lift and drag forces via the integration of the Cauchy stress tensor on the solid surface, which is difficult with existing convolution-based works. In our future work, we plan to extend the application to fluid-structure interaction systems with other deformable structures, as well as applications in 3D.

\begin{acknowledgments}
The authors would like to acknowledge the Natural Sciences and Engineering Research
Council of Canada (NSERC) and Seaspan Shipyards for the funding. This research was supported in part through computational resources and services provided by Advanced Research Computing at the University of British Columbia.
\end{acknowledgments}

\section*{Data Availability Statement}
The data that support the findings of this study are available from the corresponding author upon reasonable request.

\nocite{*}
\bibliography{references}

\end{document}